\documentclass[aps]{revtex4}
\usepackage{graphicx}
\usepackage{bm}
\begin{document}
\draft
\title{Extended RPA with ground-state correlations}
\author{M.Tohyama, S. Takahara}
\address{Kyorin University School of Medicine, 
Mitaka, Tokyo 181-8611, Japan}
\author{P. Schuck}
\address{Institut de Physique Nucl$\acute{e}$aire, IN2P3-CNRS, Universit$\acute{e}$ Paris-Sud, F-91406 Orsay Cedex, France}
\date{\today}
\begin{abstract}
We propose a time-independent method for finding a correlated ground state of 
an extended time-dependent Hartree-Fock theory, known as the time-dependent density-matrix theory (TDDM).
The correlated ground state is used to formulate the small amplitude limit of TDDM (STDDM) which is 
a version of extended RPA theories with ground-state correlations. To demonstrate the feasibility of 
the method, we calculate the ground state of $^{22}$O and study 
the first $2^+$ state and its two-phonon states using STDDM.
\end{abstract}
\maketitle
\section{Introduction}
The study of unstable nuclei is a subject of current experimental and theoretical interests.
Self-consistent theories such as the Hartree-Fock Bogoliubov theory (HFB)
and the quasi-particle random-phase approximation (QRPA), which have extensively been
used for stable nuclei, have also been applied to unstable nuclei, and the importance of ground-state
correlations has been demonstrated \cite{Khan2,Khan3,Matsuo}. Introducing a pairing field,
HFB and QRPA deal with pairing correlations in the framework of a mean field theory.
In contrast to HFB and QRPA,
the time-dependent density-matrix theory (TDDM) \cite{GT90}, which is one of extended time-dependent Hartree-Fock
theories, deals with ground-state correlations as genuine two-body correlations.
TDDM has been applied to giant resonances in stable nuclei \cite{Fabio,T99} and also to
low-lying collective states in unstable nuclei \cite{T01,T02}.
The small amplitude limit of TDDM (STDDM) \cite{TG89}, which is a time-independent version of TDDM, has also been used to
calculate low-lying states in an oxygen isotope \cite{TS}.
The importance of ground-state correlations has also been demonstrated
by these TDDM and STDDM calculations.
The correlated ground state used in these nuclear structure calculations, however, is an approximate one 
which is obtained using a time-dependent
method: The initial Hartree-Fock (HF) ground state is evolved in time using the equations of motion in TDDM and
a time-dependent residual interaction whose strength gradually approaches its intended value with
a time constant $\tau$. In the case of a solvable model where we can take $\tau$ quite large, the ground state
obtained has been found practically stationary and close to the exact one\cite{T94}.
However, in realistic cases where it is difficult to take sufficiently large $\tau$, 
the mixing of excited states, which causes spurious oscillations of some ground-state quantities,
is unavoidable though it is small \cite{T95}.
Therefore, it is anticipated to develop another method for finding a correlated ground state of the TDDM equations.
In this paper we propose a time-independent approach based on Newton's gradient method and demonstrate its feasibility by calculating
the ground state of $^{22}$O for which we have previously performed time-dependent calculations \cite{T01,T02}.
$^{22}$O is one of neutron-rich nuclei which attracts recent experimental and theoretical interests, and
is quite suitable for our present study: Although it is an open shell nucleus, the HF assumption 
which we use to obtain a starting ground state in the gradient method
is valid as first-order approximation, and the omission of the proton degrees of freedom may be allowed
in this explorative study
because of the proton shell closure.
The obtained ground state is used to construct the hamiltonian matrix of STDDM, and
the first $2^+$ state and its two phonon states in $^{22}$O are calculated.
The paper is organized as follows. In sect.2 the time-independent method for obtaining a correlated ground state
is presented. In sect.3 the formalism of STDDM is given.
The results of numerical calculations for the ground-state, the first $2^+$ state, and its two phonon states are shown
in sect.4 and sect.5 is devoted to a summary.

\section{Method for finding a correlated ground state}
The ground state $|\Phi_0\rangle$ in TDDM is constructed so that
\begin{eqnarray}
F_1(\alpha\alpha')=\langle\Phi_0|[a^+_{\alpha}a_{\alpha'},H]|\Phi_0\rangle =0,
\label{grc1}
\end{eqnarray}
\begin{eqnarray}
F_2(\alpha_1\alpha_2\alpha_2'\alpha_1')=
\langle\Phi_0|[a^+_{\alpha_1}a^+_{\alpha_2}a_{\alpha_2'}a_{\alpha_1'},H]|\Phi_0\rangle =0,
\label{grc2}
\end{eqnarray}
are satisfied, where $H$ is the total hamiltonian consisting of the kinetic energy term
and a two-body interaction, and $[~~]$ stands for the commutation relation.
In other words, the occupation matrix $n^0_{\alpha\alpha'}=\langle\Phi_0|a^+_{\alpha'}a_{\alpha}|\Phi_0\rangle$
and the two-body correlation matrix
$C^0_{\alpha_1\alpha_2\alpha_1'\alpha_2'}=\langle\Phi_0|a^+_{\alpha_1'}a^+_{\alpha_2'}a_{\alpha_2}a_{\alpha_1}|\Phi_0\rangle
-{\cal A}(n^0_{\alpha_1\alpha_1'}n^0_{\alpha_2\alpha_2'})$,
where ${\cal A}$ is the antisymmetrization operator,  
are determined so that Eqs.(\ref{grc1}) and (\ref{grc2}) are satisfied.
The expressions for Eqs.(\ref{grc1}) and (\ref{grc2}) have already been given in Ref.\cite{TG89} 
but are shown again in Appendix A.
The single-particle wavefunction $\psi_{\alpha}$ is chosen to be an eigenstate of the mean field hamiltonian $h_0(\rho_0)$:
\begin{eqnarray}
h_0(\rho_0)\psi_{\alpha}(1)=-\frac{\hbar^2\nabla^2}{2m}\psi_{\alpha}(1)+\int d2 v(1,2)
[\rho_0(2,2)\psi_{\alpha}(1)-\rho_0(1,2)\psi_{\alpha}(2)]=\epsilon_{\alpha}\psi_{\alpha}(1),
\label{hf}
\end{eqnarray}
where the numbers denote space, spin, and isospin coordinates, and the one-body density matrix $\rho_0$ is given as
\begin{eqnarray}
\rho_0(11')&=&\sum_{\alpha\alpha'}n^0_{\alpha\alpha'}\psi_{\alpha}(1)
\psi_{\alpha'}^{*}(1').
\end{eqnarray}
Attempts have been made to find a solution of Eqs.(\ref{grc1}) and (\ref{grc2}) \cite{TSW}. However, this is not evident 
partly because Eqs.(\ref{grc1}) and (\ref{grc2}) are not in the form of an eigenvalue problem. The time-dependent method
has been developed and tested for a solvable model \cite{T94} and realistic nuclei \cite{T01,T02,T95} as mentioned above. In the 
following we propose a time-independent approach using iterative Newton's gradient method.
We start from the HF ground state where $n^0_{\alpha\alpha'}=\delta_{\alpha\alpha'}~(0)$ 
for occupied (unoccupied) single-particle states
and $C^0_{\alpha_1\alpha_2\alpha_1'\alpha_2'}=0$. Then we iterate
\begin{eqnarray}
\left(
\begin{array}{c}
n^0(N+1)\\
C^0(N+1)
\end{array}
\right)
&=&
\left(
\begin{array}{c}
n^0(N)\\
C^0(N)
\end{array}
\right)-\alpha
\left(
\begin{array}{cc}
\delta F_1/\delta n^0 & \delta F_1/\delta C^0 \\
\delta F_2/\delta n^0 & \delta F_2/\delta C^0
\end{array}
\right)^{-1}
\left(
\begin{array}{c}
F_1(N)\\
F_2(N)
\end{array}
\right)
\nonumber \\
&=&\left(
\begin{array}{c}
n^0(N)\\
C^0(N)
\end{array}
\right)-\alpha
\left(
\begin{array}{cc}
a & c \\
b & d
\end{array}
\right)^{-1}
\left(
\begin{array}{c}
F_1(N)\\
F_2(N)
\end{array}
\right)
\label{iteration}
\end{eqnarray}
until convergence is achieved, where the matrix elements $a,~b,~c$, and $d$,
which also depend on the iteration step $N$, are
equivalent to those appearing in the hamiltonian matrix of STDDM. They are given in Ref.\cite{TS} and also shown in Appendix B.
We have to introduce a small parameter $\alpha$ to control the convergence process. We have tested this iterative method 
for a solvable model \cite{Lip} and found that the obtained result is equivalent to the solution  which had been obtained using the
time dependent approach \cite{T94}.

\section{Small amplitude limit of TDDM}
TDDM gives the time-evolution of
the one-body density-matrix $\rho$ and the correlated part $C_2$ 
of a two-body density-matrix \cite{GT90,WC}, and
STDDM has been formulated by 
linearizing the equations of motion for $\rho$ and $C_2$ \cite{TG89}.
The equations of STDDM for the one-body amplitude $x_{\alpha\alpha'}(\mu)$ and 
the two-body amplitude $X_{\alpha_1\alpha_2\alpha_1'\alpha_2'}(\mu)$ 
can be written in matrix form \cite{TS}
\begin{eqnarray}
\left(
\begin{array}{cc}
a & c \\
b & d
\end{array}
\right)
\left(
\begin{array}{c}
x \\
X
\end{array}
\right)
=\omega_\mu
\left(
\begin{array}{c}
x \\
X
\end{array}
\right).
\label{stddm1}
\end{eqnarray}
Eq.(\ref{stddm1}) can also be obtained from the following equations:
\begin{eqnarray}
\langle\Phi_0|[a^+_{\alpha'}a_{\alpha},H]|\Phi\rangle
&=&\omega_\mu\langle\Phi_0|a^+_{\alpha'}a_{\alpha}|\Phi\rangle ,
\label{var1}
\\
\langle\Phi_0|[a^+_{\alpha_1'}a^+_{\alpha_2'}a_{\alpha_2}a_{\alpha_1},H]|\Phi\rangle
&=&\omega_\mu\langle\Phi_0|a^+_{\alpha_1'}a^+_{\alpha_2'}a_{\alpha_2}a_{\alpha_1}|\Phi\rangle,
\label{var2}
\end{eqnarray}
where 
$|\Phi\rangle$ is the wavefunction for an excited state with excitation energy $\omega_\mu$.
Linearizing Eqs.(\ref{var1}) and (\ref{var2})
with respect to $x_{\alpha\alpha'}=\langle\Phi_0|a^+_{\alpha'}a_{\alpha}|\Phi\rangle$
and $X_{\alpha_1\alpha_2\alpha_1'\alpha_2'}=
\langle\Phi_0|a^+_{\alpha_1'}a^+_{\alpha_2'}a_{\alpha_2}a_{\alpha_1}|\Phi\rangle$, and
using
$n^0_{\alpha\alpha'}=\langle\Phi_0|a^+_{\alpha'}a_\alpha|\Phi_0\rangle$ and
$C^0_{\alpha_1\alpha_2\alpha_1'\alpha_2'}=
\langle\Phi_0|a^+_{\alpha_1'}a^+_{\alpha_2'}a_{\alpha_2}a_{\alpha_1}|\Phi_0\rangle
-{\cal A}(n^0_{\alpha_1\alpha_1'}n^0_{\alpha_2\alpha_2'})$,
we can obtain Eq.(\ref{stddm1}). The fact that the linearization of Eqs.(\ref{var1}) and (\ref{var2}) gives
Eq.(\ref{stddm1}) might explain why the matrices $a$, $b$, $c$, and $d$ 
are given by the variation
of $F_1$ and $F_2$.
The hamiltonian matrix of Eq.(\ref{stddm1}) is not hermitian as easily be understood from its
explicit form (see Appendix B). In the case of a non-hermitian hamiltonian matrix,
the ortho-normal and completeness relations are given not by the hermitian conjugate of
$|\mu\rangle=(x,~X)$ but by the left-hand-side eigenvectors $|\tilde{\mu}\rangle=(\tilde{x},~\tilde{X})$
which satisfy
\begin{eqnarray}
(\tilde{x}^*,~\tilde{X}^*)
\left(
\begin{array}{cc}
a & c \\
b & d
\end{array}
\right)
=\omega_\mu
(\tilde{x}^*,~\tilde{X}^*),
\end{eqnarray}
as explained in Ref.\cite{TS}.
When the ground-state $|\Phi_0\rangle$ is assumed to be the HF one and only the particle (p) - hole (h) and
2p - 2h amplitudes (and their complex conjugates) are taken in Eq.(\ref{stddm1}), STDDM reduces to the second RPA (SRPA)
\cite{Saw,Yan,Wam}.

The strength function is defined as
\begin{eqnarray}
S(E)&=&\sum_{E_\mu>0}|\langle\Psi_\mu|\hat{Q}|\Psi_0\rangle|^2\delta(E-E_\mu),
\end{eqnarray}
where $|\Psi_0\rangle$ is the ground-state, $|\Psi_\mu\rangle$ is an excited state 
with an excitation energy $E_\mu$, and $\hat{Q}$ an excitation operator.
The strength function in STDDM for a one-body operator $\hat{Q}_1$ is given as \cite{TS}
\begin{eqnarray}
S(E)&=-&\frac{1}{\pi}Im\{
\sum_{ Re(\omega_\mu)>0}[\left(\sum_{\alpha\alpha'}\langle\alpha|Q_1|\alpha'\rangle x_{\alpha'\alpha}(\mu)\right)
\left(\sum_{\beta\beta'}\langle\beta|Q_1|\beta'\rangle\tilde{x}^t_{\beta'\beta}(\mu)\right)^*
\frac{1}{E-\omega_\mu+i\Gamma/2}
\nonumber \\
&-&
\left(\sum_{\alpha\alpha'}\langle\alpha|Q_1|\alpha'\rangle x_{\alpha'\alpha}(\mu)\right)^*
\left(\sum_{\beta\beta'}\langle\beta|Q_1|\beta'\rangle\tilde{x}^t_{\beta'\beta}(\mu)\right)
\frac{1}{E+\omega_\mu^*+i\Gamma/2}
]\},
\label{se1}
\end{eqnarray}
where an artificial width $\Gamma$ is put to obtain a smooth distribution for $S(E)$ and
$\tilde{x}^t_{\alpha\alpha'}(\mu)$ is defined as
\begin{eqnarray}
\tilde{x}^t_{\alpha\alpha'}(\mu)
=\sum_{\lambda\lambda'}S_{11}(\alpha\alpha':\lambda\lambda')
\tilde{x}_{\lambda\lambda'}(\mu)
+\sum_{\lambda_1\lambda_2\lambda'_1\lambda'_2}
T_{12}(\alpha\alpha':\lambda_1\lambda_2\lambda_1'\lambda_2')
\tilde{X}_{\lambda_1\lambda_2\lambda'_1\lambda'_2}(\mu).
\end{eqnarray}
Here $S_{11}$ and $T_{12}$ are defined as
\begin{eqnarray}
S_{11}(\alpha\alpha':\lambda\lambda')
&=&\langle\Phi_0|[a^+_{\alpha'}a_{\alpha},a^+_{\lambda}a_{\lambda'}]|\Phi_0\rangle,
\label{S1}
\\
T_{12}(\alpha\alpha':\lambda_1\lambda_2\lambda_1'\lambda_2')
&=&\langle\Phi_0|[
a^+_{\alpha'}a_{\alpha},:a^+_{\lambda_1}a^+_{\lambda_2}a_{\lambda_2'}a_{\lambda_1'}:]|\Phi_0\rangle,
\end{eqnarray}
where $:~:$ means that 
$:a^+_{\lambda_1}a^+_{\lambda_2}a_{\lambda_2'}a_{\lambda_1'}:=a^+_{\lambda_1}a^+_{\lambda_2}a_{\lambda_2'}a_{\lambda_1'}
-{\cal A}(a^+_{\lambda_1}a_{\lambda_1'}\langle\Phi_0|a^+_{\lambda_2}a_{\lambda_2'}|\Phi_0\rangle
+a^+_{\lambda_2}a_{\lambda_2'}\langle\Phi_0|a^+_{\lambda_1}a_{\lambda_1'}|\Phi_0\rangle)$.
Similarly,
the strength function in STDDM for a two-body excitation operator $\hat{Q}_2$ is given as 
\begin{eqnarray}
S(E)&=-&\frac{1}{\pi}Im\{
\sum_{ Re(\omega_\mu)>0}[\left(\sum_{\alpha_1\alpha_2\alpha_1'\alpha_2'}
\langle\alpha_1\alpha_2|Q_2|\alpha_1'\alpha_2'\rangle X_{\alpha_1'\alpha_2'\alpha_1\alpha_2}(\mu)\right)
\nonumber \\
&\times&
\left(\sum_{\beta_1\beta_2\beta_1'\beta_2'}
\langle\beta_1\beta_2|Q_2|\beta_1'\beta_2'\rangle\tilde{X}^t_{\beta_1'\beta_2'\beta_1\beta_2}(\mu)\right)^*
\frac{1}{E-\omega_\mu+i\Gamma/2}
\nonumber \\
&-&
\left(\sum_{\alpha_1\alpha_2\alpha_1'\alpha_2'}
\langle\alpha_1\alpha_2|Q_2|\alpha_1'\alpha_2'\rangle X_{\alpha_1'\alpha_2'\alpha_1\alpha_2}(\mu)\right)^*
\nonumber \\
&\times&
\left(\sum_{\beta_1\beta_2\beta_1'\beta_2'}
\langle\beta_1\beta_2|Q_2|\beta_1'\beta_2'\rangle\tilde{X}^t_{\beta_1'\beta_2'\beta_1\beta_2}(\mu)\right)
\frac{1}{E+\omega_\mu^*+i\Gamma/2}
]\},
\label{se2}
\end{eqnarray}
where
$\tilde{X}^t_{\alpha_1\alpha_2\alpha_1\alpha_2'}(\mu)$ is given by
\begin{eqnarray}
\tilde{X}^t_{\alpha_1\alpha_2\alpha_1\alpha_2'}(\mu)
&=&\sum_{\lambda\lambda'}T_{21}(\alpha_1\alpha_2\alpha_1'\alpha_2':\lambda\lambda')
\tilde{x}_{\lambda\lambda'}(\mu)
\nonumber \\
&+&\sum_{\lambda_1\lambda_2\lambda'_1\lambda'_2}
S_{22}(\alpha_1\alpha_2\alpha_1'\alpha_2':\lambda_1\lambda_2\lambda_1'\lambda_2')
\tilde{X}_{\lambda_1\lambda_2\lambda'_1\lambda'_2}(\mu).
\end{eqnarray}
Here $T_{21}$ and $S_{22}$ are defined as 
\begin{eqnarray}
T_{21}(\alpha_1\alpha_2\alpha_1'\alpha_2':\lambda\lambda')
&=&\langle\Phi_0|[
:a^+_{\alpha_1'}a^+_{\alpha_2'}a_{\alpha_2}a_{\alpha_1}:,a^+_{\lambda}a_{\lambda'}]|\Phi_0\rangle.
\label{T2}
\\
S_{22}(\alpha_1\alpha_2\alpha_1'\alpha_2':\lambda_1\lambda_2\lambda_1'\lambda_2')
&=&\langle\Phi_0|[
:a^+_{\alpha_1'}a^+_{\alpha_2'}a_{\alpha_2}a_{\alpha_1}:,:a^+_{\lambda_1}a^+_{\lambda_2}a_{\lambda_2'}a_{\lambda_1'}:]|\Phi_0\rangle.
\end{eqnarray}
$S_{11}$, $S_{22}$, $T_{12}$ and $T_{21}$ are written in terms of $n^0_{\alpha\alpha'}$ and $C^0_{\alpha\beta\alpha'\beta'}$,
and are given explicitly in Ref.\cite{TS1}.
The strength functions in STDDM are not guaranteed to be positive definite, as is easily understood
from Eqs.(\ref{se1}) and (\ref{se2}). In practical applications this has not caused problems.
 
\section{Numerical solutions}
\subsection{Correlated ground state}
In this subsection we present how the correlated ground state of $^{22}$O is calculated and discuss 
some properties of the obtained ground state. To prepare the starting ground state for Eq.(\ref{iteration}), we perform
a static HF calculation as the first step. 
The Skyrme III (SKIII) is used as the effective interaction to generate a mean field. It has often been used as one of the
standard parameterizations of the Skyrme force in nuclear structure calculations 
even for very neutron rich nuclei \cite{Otsu,Khan2}.
We assume in the HF calculation that $1d_{5/2}$ is the last fully occupied neutron orbit of $^{22}$O.
Since Eq.(\ref{iteration}) involves matrix inversion at each iteration step, it is difficult to use a large 
number of single-particle states to solve Eq.(\ref{iteration}). In this explorative study
we only use the two neutron orbits, $1d_{5/2}$ and $2s_{1/2}$, to evolve $n^0_{\alpha\alpha'}$ and $C^0_{\alpha\beta\alpha'\beta'}$. 
Configurations consisting of these two orbits are considered to be the major components of the correlated ground state and the first $2^+$ state.
The single-particle wavefunctions are confined to a sphere
with radius 12 fm. The mesh size used is 0.1 fm. At each iteration step of Eq.(\ref{iteration}), Eq.(\ref{hf})
is solved. This means that the single-particle states also evolve in a consistent way.
In a self-consistent calculation the residual interaction should be the same as 
that used to generate the mean field. However, we had found in previous studies for giant resonances 
\cite{Toh98} that SKIII induces negligible ground-state correlations 
when the single-particle space is significantly truncated. This is the case also in this study of $^{22}$O
as will be shown below.
Therefore,
we present the results using the following pairing-type residual interaction of the density-dependent $\delta$ 
function form \cite{Chas} which is known to induce significant ground-state correlations \cite{T02}
\begin{eqnarray}
v(\bm{r}-\bm{r}')=v_{0}(1-\rho (\bm{r})/\rho_0)
\delta^3(\bm{r}-\bm{r}'),
\label{delta}
\end{eqnarray}
where $\rho (\bm{r})$ is the 
nuclear density. The parameters $\rho_0$ and $v_{0}$ are set to be 0.16fm$^{-3}$ and $-900$ MeV fm$^3$,
respectively. Similar values of $\rho_0$ and $v_{0}$ have been used in HFB
calculations \cite{Terasaki,Duguet,Yamagami} in truncated single-particle space.
The converged result somewhat depends on how the iteration process is chosen.
We found that to achieve real convergence
it is necessary to start with a small value of $v_{0}$ and gradually increase it: In the calculation shown below
we start with $v_0/20$ and increase it by $v_0/20$  for each 100 iterations.
The sum of the absolute values of the matrix elements of $F_1$ and $F_2$ is shown in Fig.1 as a function of the number of iterations.
The final value of $\sum(|F_1|+|F_2|)$ at 2100th step is $2.5\times10^{-5}$MeV.
Figure 1 demonstrates that the iteration method Eq.(\ref{iteration}) works quite well.                                              
The final single-particle energies of the $1d_{5/2}$ and $2s_{1/2}$ are -6.9 MeV and -2.8 MeV,
respectively. Their occupation probabilities are 0.98 and 0.05, respectively.  The correlation energy $E_{\rm cor}$ 
defined as 
\begin{eqnarray}
E_{\rm cor}=\frac{1}{2}\sum_{\alpha\beta\alpha'\beta'}\langle\alpha\beta|v|\alpha'\beta'\rangle C_{\alpha'\beta'\alpha\beta}
\end{eqnarray}
is -0.60 MeV.
The total energy is decreased by 0.22 MeV from the HF value: The correlation energy is largely
compensated by the increase of the HF energy.
\begin{figure}
  \begin{center}
    \includegraphics[height=7cm]{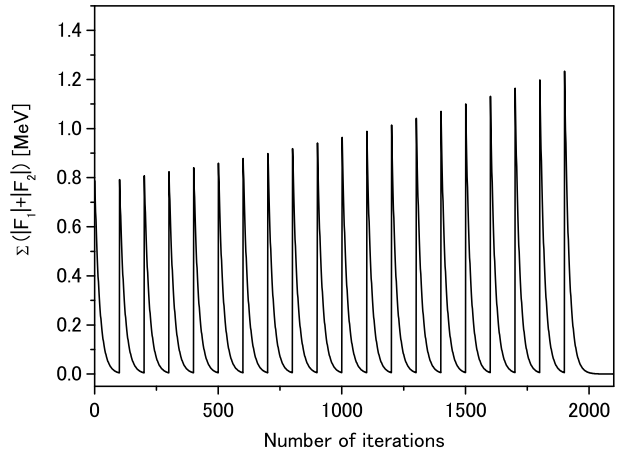}
  \end{center}
  \caption{Sum of the absolute values of the matrix elements of $F_1$ and $F_2$ as a function of the number of iterations.
Each spike corresponds to an increase of the interaction strength by $v_0/20$ (see text).}
\end{figure}
                                                                            
\subsection{Low-lying $2^+$ states}
Using the correlated ground state shown above, we solve Eq.(\ref{stddm1}) and calculate the strength function for $2^+$ states
according to Eq.(\ref{se1}). To be consistent with the calculation of $n^0_{\alpha\alpha'}$ and $C^0_{\alpha\beta\alpha'\beta'}$,
we only use the neutron $1d_{5/2}$ and $2s_{1/2}$ orbits for $x_{\alpha\alpha'}$ and $X_{\alpha\beta\alpha'\beta'}$. 
The eigenvalues of some $2^+$ states become imaginary because the hamiltonian matrix of Eq.(\ref{stddm1}) is 
not hermitian. However, their imaginary parts are quite small (less than 0.05MeV). 
Some $2^+$ states have also negative quadrupole strengths because positivity of $S(E)$ is not guaranteed.
However, the negative contributions are so small that $S(E)$ becomes positive in the entire energy region 
when it is smoothed with $\Gamma=0.5$MeV.
The obtained result for $Q_1=r^2Y_{20}$ in STDDM (solid line) is shown in Fig.2, 
where the strength functions in RPA (dot-dashed line) and SRPA (thin dotted line) is also
presented for comparison. The $2^+$ state calculated in STDDM
is energetically shifted upward and becomes significantly more collective 
as compared with that in RPA. The increase in the excitation energy is due to the lowering of the ground state, which is
realized by the increase in the unperturbed p-h energy through the coupling to $C^0_{\alpha'\beta'\alpha\beta}$. 
We will discuss this point in more detail below.
The enhancement of the collectivity of the $2^+$ state is due to
the mixing of two-body configurations. 
These properties of the first $2^+$ state under the influence of ground-state correlations
are similar to those obtained from QRPA calculations \cite{Khan3,Matsuo}. 
The reduced transition probability obtained is 68 fm$^4$. If the neutron effective charge is assumed to be $0.5e$, the value
of $B(E2:0^+\rightarrow 2_1^+)$ becomes 17 $e^2$fm$^4$, which might be comparable with the experimental value of $21\pm8~e^2$fm$^4$
\cite{Thirof}. The $2^+$ state in SRPA is simply shifted downwards due to the coupling to 2p-2h configurations
located around 8 MeV.
\begin{figure}
  \begin{center}
    \includegraphics[height=7cm]{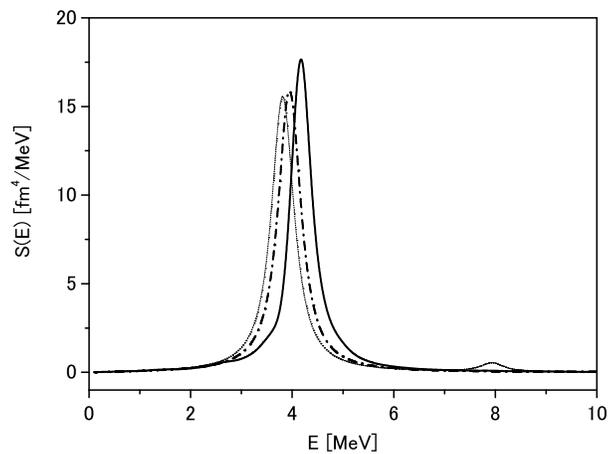}
  \end{center}
  \caption{Strength distributions of the neutron quadrupole modes
in $^{22}$O
calculated in STDDM (solid line), SRPA (thin dotted line) and RPA (dot-dashed line).
The strength functions are smoothed with $\Gamma=0.5$ MeV.}
\end{figure}

\subsection{Two phonon states}

The strength functions for the two-body operators 
$Q_2=[r^2Y_2\otimes r^2Y_2]^J_0$ with $J=$0, 2, and 4 are also calculated using Eq.(\ref{se2}) and the obtained results
are shown in Fig.3. The transition strength decreases with increasing $J$,
whereas it would be independent of $J$ if a pure phonon picture were valid.
The ratios of the transition strengths of the $2^+$ and $4^+$ states to that of the $0^+$ state are 0.55 and 0.30, respectively.
These ratios should be compared with 4/5 and 1/3, respectively, which are obtained assuming unperturbed 
two $2s_{1/2}$ particle - two $1d_{5/2}$ hole configurations.
\begin{figure}
  \begin{center}
    \includegraphics[height=7cm]{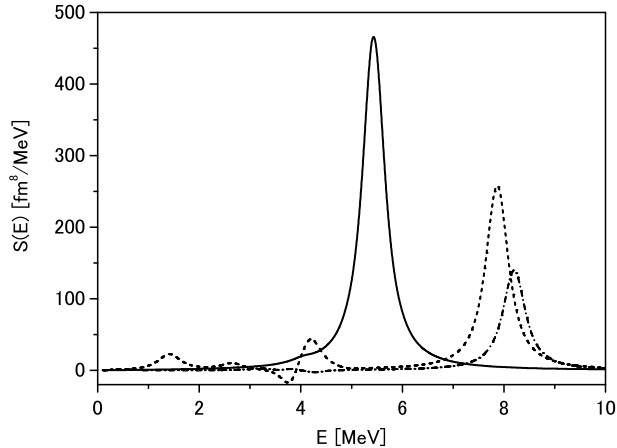}
  \end{center}
  \caption{Strength distributions of the two-phonon states of the neutron quadrupole modes
in $^{22}$O
calculated in STDDM. The solid, dotted and dot-dashed lines depict the results for $0^+,~2^+$ and $4^+$ states, respectively.
The strength functions are smoothed with $\Gamma=0.5$ MeV.}
\label{024}
\end{figure}
Although the excitation energies of the $2^+$and $4^+$ states are about twice the excitation of the 1 phonon state, 
the $0^+$ state is located at quite low energy. This originates in the nature of 
the residual interaction of the $\delta$-function form (Eq.(\ref{delta}))
which strongly favors a nucleon pair with $J^{\pi}=0^+$. The large spacing between the $0^+$ state and the $2^+$ state
is similar to that obtained from
the shell model calculation \cite{Brown}.
The strength distribution of the $2^+$ state has some small components in the low energy region (below 5 MeV).
This problem will be
discussed in the next subsection.
The transition probability between the one-phonon state and one of the two-phonon states  
is not well defined in STDDM but may be calculated as \cite{Sakata}
\begin{eqnarray}
\frac{\langle\Phi_0|[{\cal O_\mu},\hat{Q}_1]|\mu'\rangle\langle\Phi_0|[{\cal O_{\mu'}},\hat{Q}_1]|\mu\rangle}
{\langle\tilde{\mu}|\mu\rangle\langle\tilde{\mu}'|\mu'\rangle},
\label{trans}
\end{eqnarray}
where ${\cal O_\mu}$ is defined as
\begin{eqnarray}
{\cal O_\mu}=\sum (\tilde{x}^*_{\lambda\lambda'}(\mu)a^+_{\lambda'}a_\lambda
+\tilde{X}^*_{\lambda_1\lambda_2\lambda_1'\lambda_2'}(\mu)
a^+_{\lambda_1'}a^+_{\lambda_2'}a_{\lambda_2}a_{\lambda_1}).
\end{eqnarray}
The reduced transition probabilities from the $0^+$, $2^+$, and $4^+$ states to the 1 phonon state
are 16, 14, and 12 fm$^4$, respectively, which are close to the
value 68/5$\approx$14 fm$^4$ for the transition  from the firs $2^+$ state to the ground state. 
If a pure phonon picture were valid, these values would be independent of the angular momenta
of the 2 phonon states.

In the following we compare STDDM with SRPA.
\begin{figure}
  \begin{center}
    \includegraphics[height=7cm]{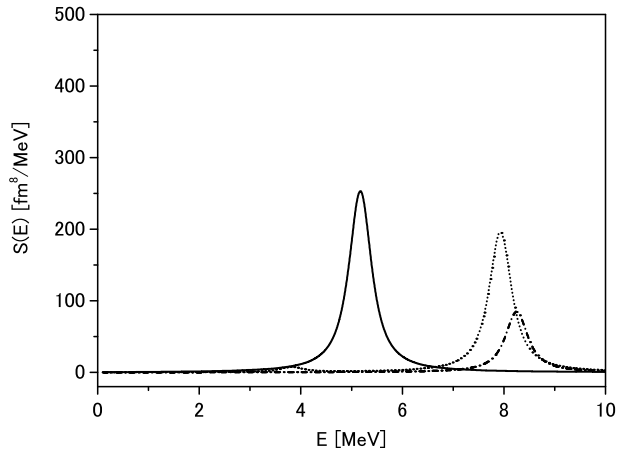}
  \end{center}
  \caption{Strength distributions of the two-phonon states of the neutron quadrupole mode
in $^{22}$O
calculated in SRPA. The solid, dotted and dot-dashed lines depict the results for $0^+,~2^+$ and $4^+$ states, respectively.
The strength functions are smoothed with $\Gamma=0.5$ MeV.}
\end{figure}
The strength functions calculated in SRPA are shown in Fig. 4. The transition strengths in SRPA 
are much smaller than those in STDDM. Note that Fig. 4 is drawn in the same scale as Fig. 3.
It had been pointed out \cite{Sakata} that the $X_{phph}$ amplitude is important to reproduce the collectivity of 
low-lying two phonon states. 
To investigate this point, we performed a calculation for the $0^+$ state
using a modified SRPA (mSRPA) which includes $X_{phph}$: mSRPA can be obtained from Eq.(6) by evaluating the hamiltonian matrix
using the HF ground state. The result is shown in Fig. 5.
The $X_{phph}$ amplitude significantly enhances the collectivity of the  $0^+$ state. However, the transition strength in mSRPA
is still smaller that that in STDDM. In Fig. 5 we also show the result of a further modified SRPA
which includes all two-body amplitudes except for the 3p-1h, 1h-3p, 3h-1p and
1p-3h amplitudes. The 3p-1h and 3h-1p amplitudes do not increase the transition strength, although they
change somewhat the form of the strength function. Now the transition strength becomes 91\% of the strength in STDDM.
Thus all two-body amplitudes except for the 3p-1h and 3h-1p types seem to be important to describe the collectivity of the 
low-lying 2 phonon states.
This is in contrast to two-phonon states of giant resonances where the transition strengths are exhausted by the 2p-2h, 2h-2p
and 1p1h-1p1h amplitudes \cite{DGQR}.
The fact that the $0^+$ state in STDDM is not much lowered as compared with the modified SRPA results
is due to ground-state correlations.
\begin{figure}
  \begin{center}
    \includegraphics[height=7cm]{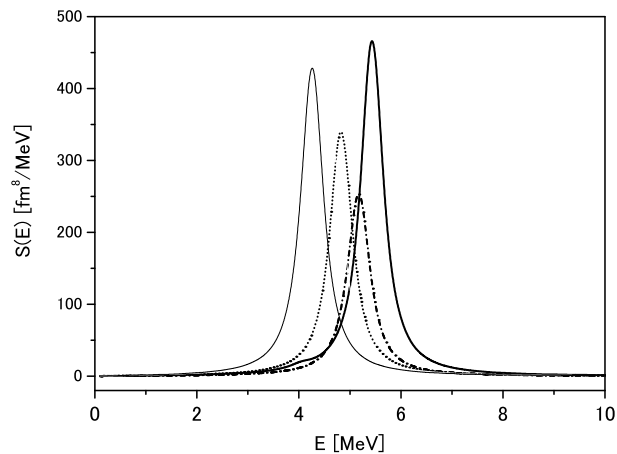}
  \end{center}
  \caption{Strength distributions of the $0^+$ two-phonon state of the neutron quadrupole modes
in $^{22}$O
calculated in STDDM (solid line), SRPA (dot-dashed line) and mSRPA (dotted line) which includes $X_{phph}$ type amplitudes. 
The thin solid line indicates
the result calculated using further modified SRPA where the $X_{pppp}$ and $X_{hhhh}$ type amplitudes are added to mSRPA.
The strength functions are smoothed with $\Gamma=0.5$ MeV.}
\end{figure}

We also performed a self-consistent STDDM calculation using
SKIII as a residual interaction. For simplicity, we neglected the spin-orbit force.
Since there is a strong cancellation between the momentum dependent and
independent terms,
SKIII induces quite weak ground-state correlations in the truncated single-particle 
space considered in this work: The occupation probabilities of
the $1d_{5/2}$ and $2s_{1/2}$ are 0.9999 and 0.0003, respectively.
The strength functions for the two-phonon states 
are shown in Fig.6.  
The ratios of the g.s.$\rightarrow2^+$ and $4^+$ transition strengths 
to the g.s.$\rightarrow0^+$ one
are 0.77 and 0.33, respectively,  which are close to 4/5 and 1/3 obtained assuming the unperturbed configurations
as discussed above.
\begin{figure}
  \begin{center}
    \includegraphics[height=7cm]{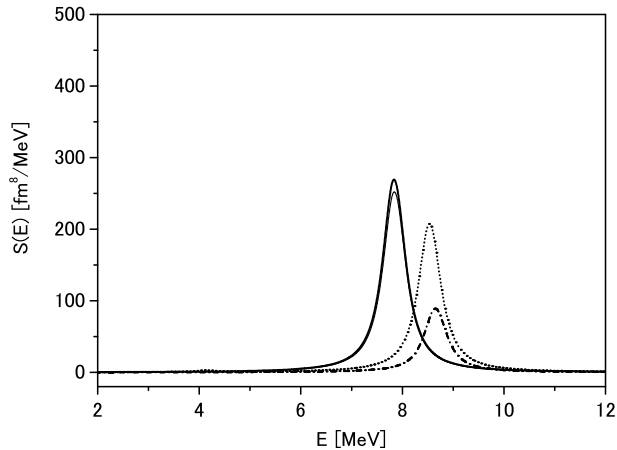}
  \end{center}
  \caption{Strength distributions of the two-phonon states of the neutron quadrupole modes
in $^{22}$O
calculated in STDDM using SKIII.
The solid, dotted and dot-dashed lines depict the results for $0^+,~2^+$ and $4^+$ states, respectively.
The result in SRPA for the $0^+$ state is also shown with the thin solid line.
The strength functions are smoothed with $\Gamma=0.5$ MeV.}
\end{figure}
Since SKIII is effectively a very weak interaction in the truncated single-particle space, 
there is no significant difference between the STDDM and SRPA results.

\subsection{Incoherent states}
\begin{figure}
  \begin{center}
    \includegraphics[height=7cm]{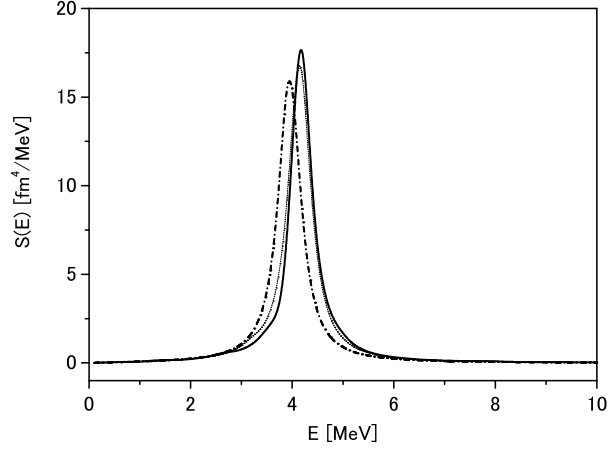}
  \end{center}
  \caption{Strength distributions of the neutron quadrupole mode
calculated in STDDM (solid line) and mSTDDM (thin dotted line). The result in RPA (dot-dashed line)
is also shown for comparison. The strength functions are smoothed with $\Gamma=0.5$ MeV.}
\label{mstddm1}
\end{figure}

\begin{figure}
  \begin{center}
    \includegraphics[height=7cm]{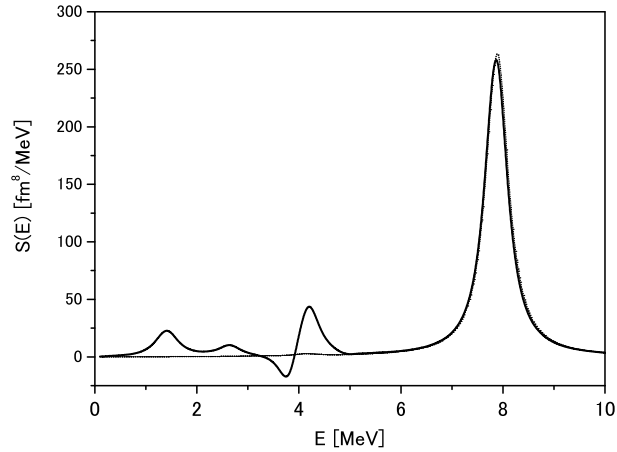}
  \end{center}
  \caption{Strength distributions of the $2^+$ two-phonon state 
calculated in STDDM (solid line) and mSTDDM (thin dotted line).The strength functions are smoothed with $\Gamma=0.5$ MeV.}
\label{mstddm2}
\end{figure}

Now we discuss incoherent states which are seen in the low energy region of 
the strength function of the 2 phonon state with $J^\pi=2^+$ (Fig. \ref{024}). These incoherent states have large components
of the 3p-1h and 1p-3h configurations which have the same unperturbed energy as the 1p-1h configurations.
Since the 3p-1h, 3h-1p amplitudes and their complex conjugates are used
as dynamical amplitudes in Eq.(\ref{stddm1}), 
these incoherent states naturally appear as eigenstates. However, the important physical role played by these
amplitudes seems to be a rather kinematical one. To clarify this point, we use a different expression for Eq.(\ref{stddm1}) \cite{TS}.
When the eigenvector $(x, X)$ in STDDM is transformed to $(y, Y)$ as
\begin{eqnarray}
\left(
\begin{array}{c}
x \\
X
\end{array}
\right)=
\left(
\begin{array}{cc}
S_{11} & T_{12} \\
T_{21}& S_{22}
\end{array}
\right)
\left(
\begin{array}{c}
y \\
Y
\end{array}
\right),
\label{xXyY}
\end{eqnarray}
Eq.(\ref{stddm1}) becomes
\begin{eqnarray}
\left(
\begin{array}{cc}
aS_{11}+cT_{21} & aT_{12}+cS_{22} \\
bS_{11}+dT_{21} & bT_{12}+dS_{22}
\end{array}
\right)
\left(
\begin{array}{c}
y \\
Y
\end{array}
\right)
=\omega_\mu
\left(
\begin{array}{cc}
S_{11} & T_{12} \\
T_{21} & S_{22}
\end{array}
\right)
\left(
\begin{array}{c}
y \\
Y
\end{array}
\right).
\label{erpa4}
\end{eqnarray}
It has been pointed out \cite{TS} that this form of STDDM is a reasonable
approximation for an extended RPA with hermiticity \cite{TS1}.
The kinematical effects of ground-state correlations are included in the hamiltonian matrix of Eq.(\ref{erpa4}).
For example, the increase of the unperturbed energy of 1 phonon states due to ground-state correlations is
given by $cT_{21}$ in $aS_{11}+cT_{21}$, whereas $aS_{11}$ 
describes the renormalization of the RPA matrix due to the change in occupation
factors. Omitting 
the 3p-1h and 3h-1p amplitudes and their complex conjugates in $Y$, we can avoid dynamical contributions of these amplitudes. This modified STDDM is referred to as mSTDDM.
The obtained results for the 1 phonon state and the 2 phonon state with $J^\pi=2^+$ are shown with thin dotted lines
in Figs.\ref{mstddm1} and \ref{mstddm2}, respectively. The coherent states are little affected by the omission
of the 3p-1h and 3h-1p amplitudes and their complex conjugates, and the incoherent states seen in the 2 phonon state 
can be eliminated.
Thus it would be better to use Eq.(\ref{erpa4}) without the 3p-1h and 3h-1p type amplitudes instead of using
Eq.(\ref{stddm1}) with full two-body amplitudes when we are interested in the calculation of the strength functions.
However, we also found that the omission of the the 3p-1h and 3h-1p type amplitudes affects
the transition probability between excited states when it is evaluated using Eq.(\ref{trans}).

Finally we discuss incoherent states associated with one-body and two-body amplitudes which have zero unperturbed energy:
$x_{pp}$, $x_{hh}$, $X_{phph}$, $X_{pppp}$, and $X_{hhhh}$ are such amplitudes.
Some incoherent solutions of Eq.(\ref{stddm1}) or Eq.(\ref{erpa4}) have finite energies but most of them stay at zero or nearly zero energies.
The transition strengths to these states are so small that they are invisible in the strength functions shown above.
If these amplitudes are neglected and only $x_{ph}$, $x_{hp}$, $X_{pphh}$, and $X_{hhpp}$ are taken in Eq.(\ref{erpa4}),
these incoherent states disappear. However, a serious problem arises as to the collectivity of coherent states, especially 
of 2 phonon states, as explained above. Therefore,  what we must do would be to interpret
these incoherent states as unphysical ones and consider only coherent states. 

\subsection{Effects of three-body amplitudes}
\begin{figure}
  \begin{center}
    \includegraphics[height=7cm]{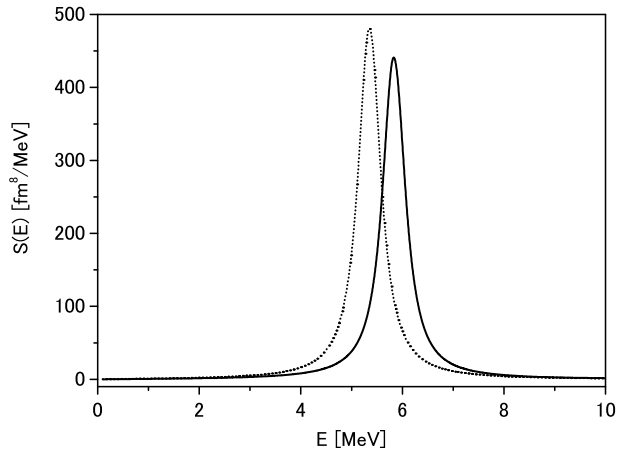}
  \end{center}
  \caption{Strength distributions of the $0^+$ two-phonon state 
calculated in mSTDDM with Eq.(\ref{3-body}) (solid line) and without it (dotted line).
The strength functions are smoothed with $\Gamma=0.5$ MeV.}
\label{mstddm3}
\end{figure}

The hamiltonian matrix $bT_{12}+dS_{22}$ for the two-body amplitudes in Eq.(\ref{erpa4}) cannot represent all terms
in the extended RPA of Ref.\cite{TS1}. The terms coming from
\begin{eqnarray}
\sum_{\gamma_1\gamma_2\gamma_3\gamma_1'\gamma_2'\gamma_3'}
e(\alpha_1\alpha_2\alpha_1'\alpha_2':\gamma_1\gamma_2\gamma_3\gamma_1'\gamma_2'\gamma_3')
T_{32}(\gamma_1\gamma_2\gamma_3\gamma_1'\gamma_2'\gamma_3':\lambda_1\lambda_2\lambda_1'\lambda_2')
\label{3-body}
\end{eqnarray}
are the missing terms.
Here $e$ is a matrix that would come into the hamiltonian matrix if we included a three-body amplitude in 
the evaluation of the left-hand side of Eq.(\ref{var2}), and
is given as
\begin{eqnarray}
e(\alpha_1\alpha_2\alpha_1'\alpha_2':\gamma_1\gamma_2\gamma_3\gamma_1'\gamma_2'\gamma_3')&=&
-\langle\alpha_1\gamma_3'|v|\gamma_1\gamma_2\rangle
\delta_{\alpha_2\gamma_3}\delta_{\alpha_1'\gamma_1'}\delta_{\alpha_2'\gamma_2'}
+\langle\alpha_2\gamma_3'|v|\gamma_1\gamma_2\rangle
\delta_{\alpha_1\gamma_3}\delta_{\alpha_1'\gamma_1'}\delta_{\alpha_2'\gamma_2'}
\nonumber \\
&+&\langle\gamma_1'\gamma_2'|v|\alpha_1'\gamma_3\rangle
\delta_{\alpha_1\gamma_1}\delta_{\alpha_2\gamma_2}\delta_{\alpha_2'\gamma_3'}
-\langle\gamma_1'\gamma_2'|v|\alpha_2'\gamma_3\rangle
\delta_{\alpha_1\gamma_1}\delta_{\alpha_2\gamma_2}\delta_{\alpha_1'\gamma_3'}.
\end{eqnarray}
$T_{32}$ is defined as
\begin{eqnarray}
T_{32}(\gamma_1\gamma_2\gamma_3\gamma_1'\gamma_2'\gamma_3':\lambda_1\lambda_2\lambda_1'\lambda_2')
=\langle\Phi_0|[:a^+_{\gamma_1'}a^+_{\gamma_2'}a^+_{\gamma_3'}a_{\gamma_3}a_{\gamma_2}a_{\gamma_1}:,
:a^+_{\lambda_1}a^+_{\lambda_2}a_{\lambda_2'}a_{\lambda_1'}]|\Phi_0\rangle.
\end{eqnarray}
This $eT_{32}$ term contributes to the two-body sector of the hamiltonian matrix in Eq.(\ref{erpa4})
when the transformation of Eq.(\ref{xXyY}) includes the three-body amplitudes.
Kinematical effects of ground-state correlations such as the increase in unperturbed energies of 2 phonon states
are expressed by some terms in Eq.(\ref{3-body}). We performed a calculation in mSTDDM including Eq.(\ref{3-body})
and found that the 1 phonon state is little affected: The excitation energy of the first $2^+$ state is unchanged
and the transition strength to the first $2^+$ state is slightly increased (by 2.5\%). The inclusion of the terms
in Eq.(\ref{3-body}) somewhat affects the properties of 
the 2 phonon states as expected: The excitation energies are increased by $0.2\sim0.5$ MeV and the transition strengths
are decreased by $8\sim20$ \%, depending on $J$ of the 2 phonon states. As an example the strength functions for 
the $0^+$ state calculated in mSTDDM with 
the terms in Eq.(\ref{3-body}) (solid line) and without them (dotted line) are shown in Fig. \ref{mstddm3}.
Thus the terms in Eq.(\ref{3-body}) need to be included in quantitative study of 2 phonon states.

\section{Summary}
We proposed a time-independent method for obtaining a correlated ground-state of the time-dependent
density-matrix theory (TDDM). The method was applied to obtain the correlated ground state of $^{22}$O.
The eigenstates of the small amplitude limit of TDDM (STDDM) were calculated for the first $2^+$ state in
$^{22}$O using the correlated ground state.
It is found that STDDM properly deals with the effects of ground-state correlations
on the low-lying $2^+$ state and that the non-hermiticity of STDDM is quite moderate one:
The eigenvalues have quite small imaginary parts and the strength function is practically positive
definite although it is not guaranteed in its non-hermitian form.
The 2 phonon states of the $2^+$ state were also studied. It was found that the 2 phonon state with $J^\pi=0^+$ appears at very low
excitation energy. This originates in the nature of the zero range force used. The results 
obtained using the Skyrme III force as a residual interaction were also presented. It was found that SKIII acts as a weak residual interaction
in the very truncated single-particle space considered in this study. The physical roles played by the 3 particle -1 hole
and 3 hole - 1 particle type amplitudes in STDDM were discussed and a method for eliminating the incoherent states
associated with these amplitudes was presented. It was also pointed out that the self-energy terms for unperturbed
2 phonon configurations, which are missing in STDDM, can be included using the extended RPA formalism of Ref.\cite{TS,TS1}.

\appendix
\section{}
When $\psi_{\alpha}$ is chosen to be an eigenstate of the mean field hamiltonian (Eq.(\ref{hf})),
Eqs.(\ref{grc1}) and (\ref{grc2}) become
\begin{eqnarray}
(\epsilon_{\alpha'}-\epsilon_{\alpha})n^0_{\alpha\alpha'}&=&
\sum_{\lambda_1\lambda_2\lambda_3}(C^0_{\lambda_1\lambda_2\alpha'\lambda_3}
\langle\alpha\lambda_3|v|\lambda_1\lambda_2\rangle -C^0_{\alpha\lambda_3\lambda_1\lambda_2}
\langle\lambda_1\lambda_2|v|\alpha'\lambda_3\rangle)
\label{Agrc1}
\\
(\epsilon_{\alpha'}+\epsilon_{\beta'}-\epsilon_{\alpha}-\epsilon_{\beta})
C^0_{\alpha\beta\alpha'\beta'}&=&
B^0_{\alpha\beta\alpha'\beta'}+P^0_{\alpha\beta\alpha'\beta'}+H^0_{\alpha\beta\alpha'\beta'},
\label{Agrc2}
\end{eqnarray}
where
\begin{eqnarray}
B^0_{\alpha\beta\alpha'\beta'}&=&\sum_{\lambda_1\lambda_2\lambda_3\lambda_4}
\langle\lambda_1\lambda_2|v|\lambda_3\lambda_4\rangle_A
[(\delta_{\alpha\lambda_1}-n^0_{\alpha\lambda_1})(\delta_{\beta\lambda_2}-n^0_{\beta\lambda_2})
n^0_{\lambda_3\alpha'}n^0_{\lambda_4\beta'}
\nonumber \\
&-&n^0_{\alpha\lambda_1}n^0_{\beta\lambda_2}(\delta_{\lambda_3\alpha'}-n^0_{\lambda_3\alpha'})
(\delta_{\lambda_4\beta'}-n^0_{\lambda_4\beta'})],
\\
P^0_{\alpha\beta\alpha'\beta'}&=&\sum_{\lambda_1\lambda_2\lambda_3\lambda_4}
\langle\lambda_1\lambda_2|v|\lambda_3\lambda_4\rangle
[(\delta_{\alpha\lambda_1}\delta_{\beta\lambda_2}
-\delta_{\alpha\lambda_1}n^0_{\beta\lambda_2}
-n^0_{\alpha\lambda_1}\delta_{\beta\lambda_2})
C^0_{\lambda_3\lambda_4\alpha'\beta'}
\nonumber \\
&-&(\delta_{\lambda_3\alpha'}\delta_{\lambda_4\beta'}
-\delta_{\lambda_3\alpha'}n^0_{\lambda_4\beta'}
-n^0_{\lambda_3\alpha'}\delta_{\lambda_4\beta'})
C^0_{\alpha\beta\lambda_1\lambda_2}],
\\
H^0_{\alpha\beta\alpha'\beta'}&=&\sum_{\lambda_1\lambda_2\lambda_3\lambda_4}
\langle\lambda_1\lambda_2|v|\lambda_3\lambda_4\rangle_A
[\delta_{\alpha\lambda_1}(n^0_{\lambda_3\alpha'}C^0_{\lambda_4\beta\lambda_2\beta'}
-n^0_{\lambda_3\beta'}C^0_{\lambda_4\beta\lambda_2\alpha'})
\nonumber \\
&+&\delta_{\beta\lambda_2}(n^0_{\lambda_4\beta'}C^0_{\lambda_3\alpha\lambda_1\alpha'}
-n^0_{\lambda_4\alpha'}C^0_{\lambda_3\alpha\lambda_1\beta'})
\nonumber \\
&-&\delta_{\alpha'\lambda_3}(n^0_{\alpha\lambda_1}C^0_{\lambda_4\beta\lambda_2\beta'}
-n^0_{\beta\lambda_1}C^0_{\lambda_4\alpha\lambda_2\beta'})
\nonumber \\
&-&\delta_{\beta'\lambda_4}(n^0_{\beta\lambda_2}C^0_{\lambda_3\alpha\lambda_1\alpha'}
-n^0_{\alpha\lambda_2}C^0_{\lambda_3\beta\lambda_1\alpha'})].
\label{H0}
\end{eqnarray}
Here the subscript $A$ indicates that the corresponding matrix is antisymmetrized.

\section{}
The matrices in Eq.(\ref{stddm1}) are shown below:
\begin{eqnarray}
a(\alpha\alpha':\lambda\lambda')&=&(\epsilon_{\alpha}-\epsilon_{\alpha'})\delta_{\alpha\lambda}\delta_{\alpha'\lambda'}
-\sum_{\beta}(\langle\beta\lambda'|v|\alpha'\lambda\rangle_An^0_{\alpha\beta}
-\langle\alpha\lambda'|v|\beta\lambda\rangle_An^0_{\beta\alpha'}),
\\
b(\alpha_1\alpha_2\alpha_1'\alpha_2':\lambda\lambda')&=&
-\delta_{\alpha_1\lambda}\{\sum_{\beta\gamma\delta}((\delta_{\alpha_2\beta}-n^0_{\alpha_2\beta})
n^0_{\gamma\alpha_1'}n^0_{\delta\alpha_2'}
+n^0_{\alpha_2\beta}(\delta_{\gamma\alpha_1'}-n^0_{\gamma\alpha_1'})(\delta_{\delta\alpha_2'}-n^0_{\delta\alpha_2'})
\langle\lambda'\beta|v|\gamma\delta\rangle_A
\nonumber \\
&+&\sum_{\beta\gamma}(\langle\lambda'\alpha_2|v|\beta\gamma\rangle C^0_{\beta\gamma\alpha_1'\alpha_2'}
+\langle\lambda'\beta|v|\alpha_1'\gamma\rangle_A C^0_{\alpha_2\gamma\alpha_2'\beta}
-\langle\lambda'\beta|v|\alpha_2'\gamma\rangle_A C^0_{\alpha_2\gamma\alpha_1'\beta}\}
\nonumber \\
&+&\delta_{\alpha_2\lambda}\{\sum_{\beta\gamma\delta}[(\delta_{\alpha_1\beta}-n^0_{\alpha_1\beta})
n^0_{\gamma\alpha_1'}n^0_{\delta\alpha_2'}
+n^0_{\alpha_1\beta}(\delta_{\gamma\alpha_1'}-n^0_{\gamma\alpha_1'})(\delta_{\delta\alpha_2'}-n^0_{\delta\alpha_2'})
\langle\lambda'\beta|v|\gamma\delta\rangle_A]
\nonumber \\
&+&\sum_{\beta\gamma}[\langle\lambda'\alpha_1|v|\beta\gamma\rangle C^0_{\beta\gamma\alpha_1'\alpha_2'}
+\langle\lambda'\beta|v|\alpha_1'\gamma\rangle_A C^0_{\alpha_1\gamma\alpha_2'\beta}
-\langle\lambda'\beta|v|\alpha_2'\gamma\rangle_A C^0_{\alpha_1\gamma\alpha_1'\beta}]\}
\nonumber \\
&+&\delta_{\alpha_1'\lambda'}\{\sum_{\beta\gamma\delta}[(\delta_{\delta\alpha_2'}-n^0_{\delta\alpha_2'})
n^0_{\alpha_1\beta}n^0_{\alpha_2\gamma}
+n^0_{\delta\alpha_2'}
(\delta_{\alpha_1\beta}-n^0_{\alpha_1\beta})(\delta_{\alpha_2\gamma}-n^0_{\alpha_2\gamma})
\langle\beta\gamma|v|\lambda\delta\rangle_A]
\nonumber \\
&+&\sum_{\beta\gamma}[\langle\beta\gamma|v|\lambda\alpha_2'\rangle C^0_{\alpha_1\alpha_2\beta\gamma}
+\langle\alpha_1\beta|v|\lambda\gamma\rangle_A C^0_{\alpha_2\gamma\alpha_2'\beta}
-\langle\alpha_2\beta|v|\lambda\gamma\rangle_A C^0_{\alpha_1\gamma\alpha_2'\beta}]\}
\nonumber \\
&-&\delta_{\alpha_2'\lambda'}\{\sum_{\beta\gamma\delta}[(\delta_{\delta\alpha_1'}-n^0_{\delta\alpha_1'})
n^0_{\alpha_1\beta}n^0_{\alpha_2\gamma}
+n^0_{\delta\alpha_1'}
(\delta_{\alpha_1\beta}-n^0_{\alpha_1\beta})(\delta_{\alpha_2\gamma}-n^0_{\alpha_2\gamma})
\langle\beta\gamma|v|\lambda\delta\rangle_A]
\nonumber \\
&+&\sum_{\beta\gamma}[\langle\beta\gamma|v|\lambda\alpha_1'\rangle C^0_{\alpha_1\alpha_2\beta\gamma}
+\langle\alpha_1\beta|v|\lambda\gamma\rangle_A C^0_{\alpha_2\gamma\alpha_1'\beta}
-\langle\alpha_2\beta|v|\lambda\gamma\rangle_A C^0_{\alpha_1\gamma\alpha_1'\beta}]\}
\nonumber \\
&+&\sum_{\beta}[\langle\alpha_1\lambda'|v|\beta\lambda\rangle_A C^0_{\beta\alpha_2\alpha_1'\alpha_2'}
-\langle\alpha_2\lambda'|v|\beta\lambda\rangle_A C^0_{\beta\alpha_1\alpha_1'\alpha_2'}
\nonumber \\
&-&\langle\beta\lambda'|v|\alpha_2'\lambda\rangle_A C^0_{\alpha_1\alpha_2\alpha_1'\beta}
+\langle\beta\lambda'|v|\alpha_1'\lambda\rangle_A C^0_{\alpha_1\alpha_2\alpha_2'\beta}],
\\
c(\alpha\alpha':\lambda_1\lambda_2\lambda_1'\lambda_2')&=&
\langle\alpha\lambda_2'|v|\lambda_1\lambda_2\rangle\delta_{\alpha'\lambda_1'}
-\langle\lambda_1'\lambda_2'|v|\alpha'\lambda_2\rangle\delta_{\alpha\lambda_1},
\\
d(\alpha_1\alpha_2\alpha_1'\alpha_2':\lambda_1\lambda_2\lambda_1'\lambda_2')&=&
(\epsilon_{\alpha_1}+\epsilon_{\alpha_2}-\epsilon_{\alpha_1'}-\epsilon_{\alpha_2'})
\delta_{\alpha_1\lambda_1}\delta_{\alpha_2\lambda_2}
\delta_{\alpha_1'\lambda_1'}\delta_{\alpha_2'\lambda_2'}
\nonumber \\
&+&\delta_{\alpha_1'\lambda_1'}\delta_{\alpha_2'\lambda_2'}
\sum_{\beta\gamma}(\delta_{\alpha_1\beta}\delta_{\alpha_2\gamma}
-\delta_{\alpha_2\gamma}n^0_{\alpha_1\beta}
-\delta_{\alpha_1\beta}n^0_{\alpha_2\gamma})\langle\beta\gamma|v|\lambda_1\lambda_2\rangle
\nonumber \\
&-&\delta_{\alpha_1\lambda_1}\delta_{\alpha_2\lambda_2}
\sum_{\beta\gamma}(\delta_{\alpha_1'\beta}\delta_{\alpha_2'\gamma}
-\delta_{\alpha_2'\gamma}n^0_{\alpha_1'\beta}
-\delta_{\alpha_1'\beta}n^0_{\alpha_2'\gamma})
\langle\lambda_1'\lambda_2'|v|\beta\gamma\rangle
\nonumber \\
&+&\delta_{\alpha_2\lambda_2}\delta_{\alpha_2'\lambda_2'}
\sum_{\beta}(\langle\alpha_1\lambda_1'|v|\beta\lambda_1\rangle_An^0_{\beta\alpha_1'}
-\langle\beta\lambda_1'|v|\alpha_1'\lambda_1\rangle_An^0_{\alpha_1\beta})
\nonumber \\
&+&\delta_{\alpha_2\lambda_2}\delta_{\alpha_1'\lambda_1'}
\sum_{\beta}(\langle\alpha_1\lambda_2'|v|\beta\lambda_1\rangle_An^0_{\beta\alpha_2'}
-\langle\beta\lambda_2'|v|\alpha_2'\lambda_1\rangle_An^0_{\alpha_1\beta})
\nonumber \\
&+&\delta_{\alpha_1\lambda_1}\delta_{\alpha_1'\lambda_1'}
\sum_{\beta}(\langle\alpha_2\lambda_2'|v|\beta\lambda_2\rangle_An^0_{\beta\alpha_2'}
-\langle\beta\lambda_2'|v|\alpha_2'\lambda_2\rangle_An^0_{\alpha_2\beta})
\nonumber \\
&+&\delta_{\alpha_1\lambda_1}\delta_{\alpha_2'\lambda_2'}
\sum_{\beta}(\langle\alpha_2\lambda_1'|v|\beta\lambda_2\rangle_An^0_{\beta\alpha_1'}
-\langle\beta\lambda_1'|v|\alpha_1'\lambda_2\rangle_An^0_{\alpha_2\beta}),
\end{eqnarray}
where $n^0_{\alpha\alpha'}=\langle\Phi_0|a^+_{\alpha'}a_\alpha|\Phi_0\rangle$ and
$C^0_{\alpha_1\alpha_2\alpha_1'\alpha_2'}=
\langle\Phi_0|a^+_{\alpha_1'}a^+_{\alpha_2'}a_{\alpha_2}a_{\alpha_1}|\Phi_0\rangle
-{\cal A}(n^0_{\alpha_1\alpha_1'}n^0_{\alpha_2\alpha_2'})$.

\end{document}